# Circular dichroism of second harmonic generation response

S. W. Lovesey and G. van der Laan

*Diamond Light Source, Chilton, Didcot, Oxfordshire OX11 0DE, United Kingdom*

**Abstract**. Polarization-dependent photon spectroscopy (dichroism) of the second-harmonic generation response is shown to reveal chiral and magnetic properties of a sample. Two dichroic signals are allowed with electric-dipole and electric-quadrupole scattering events, and both require circular polarization in the primary beam. Explicit expressions for the signals are derived using theoretical techniques from atomic physics.

## I. INTRODUCTION

Second-harmonic generation (SHG) was first demonstrated from a quartz crystal soon after the invention of the ruby laser [1]. It is often referred to as a frequency doubling phenomenon since two photons of frequency $\omega$ are converted into one photon of frequency $2\omega$, in a process depicted in Fig. 1. The optical polarization is a quadratic function of the components of the symmetric tensor $E(\omega)E(\omega)$, where $E(\omega)$ is the optical electric field. The 18 coefficients which occur in this function are subject to restrictions due to the point symmetry of the medium. These restrictions are identical with those governing the piezoelectric coefficients [2].

The origin and nature of natural circular dichroism (NCD) were well-understood in the 1960s [3]. Chemists exploited NCD in the SHG response three decades later [4], with a surge of exploitation in studies of nanostructures and surfaces [5, 6]. As the SHG microscope is able to selectively observe the region in a sample where spatial inversion symmetry is broken, it is a valuable tool for investigating the molecular ordering and structural organization in biological samples [7]. Similar to ordinary NCD spectroscopy, signals in the SHG response are weak and measurable only near resonance. Magnetic circular dichroism (MCD) in the SHG response has also been convincingly demonstrated [8].

The recent advent of free-electron lasers (FELs) in the energy range from extreme ultraviolet to x-rays allows to explore SHG effects involving core-level resonances. As an example, Yamamoto *et al.* [9] measured SHG at the Fe 3*p* edge of gallium ferrate (GaFeO$_3$) using soft x-ray FEL radiation. Other nonlinear optical techniques observed in the extreme-ultraviolet include four-wave mixing [10] and x-ray two-photon absorption [11]. Furthermore, high brightness ultrafast x-ray pulses from an x-ray FEL provide the capability for time-resolved probing of atomic scale structure and electronic states in a material using x-ray scattering and x-ray spectroscopy

Absorption and diffraction are two sides of one coin, since they have in common the x-ray scattering length, $f$. Absorption and its dependence on photon polarization and sample properties (dichroism) are related to the scattering length with no deflection of the x-ray beam, while intensities of Bragg spots in a diffraction pattern are proportional to $|f|^2$. Third-order perturbation theory yields the contribution to $f$ that accounts for the SHG response, and the well-established result is reported in many papers, e.g., Eq. (9) in Ref. [4]. Pertinent results are gathered in an Appendix for the convenience of the reader. Like the more familiar Kramers-Heisenberg amplitude [12], $f$ is a sum of energy denominators that can vanish for specific photon energies, in the absence of uncertainty in the energies of virtual intermediate states, leading to resonant processes. Tertiary matrix-elements in (A3), subsequently referred to as $G$, fix the weight of each of three contributions to the scattering length (one more than in the Kramers-Heisenberg amplitude (A1) used for ordinary dichroism, where matrix elements are dyadic). Photon polarization resides in $G$s, and they are central characters in the investigation reported in this contribution on the theory of dichroism in the SHG response. We demonstrate that electronic operators in $G$ form spherical multipoles, which offer a perspicacious description of corresponding dichroic signals akin to that already achieved for signals from ordinary polarization-dependent photon spectroscopy [13-16].

Individual matrix elements in the Kramers-Heisenberg amplitude and $G$ are those of $\boldsymbol{\varepsilon} \cdot \mathfrak{J}(\mathbf{q})$, where $\boldsymbol{\varepsilon}$ is the photon polarization vector and $\mathfrak{J}$ a current operator that includes the photon wavevector, $\mathbf{q}$ ($\boldsymbol{\varepsilon} \cdot \mathbf{q} = 0$), and all electronic degrees of freedom [16]. Optical phenomena are accounted for by the first few terms of $\mathfrak{J}(\mathbf{q})$ in an expansion using ($q a_o$) as the small quantity, where $a_o$ is the Bohr radius, and the result is (A2). We include three different terms in (A2) labelled by their engagement of electronic degrees of freedom, the electric dipole E1 ≈ ($e a_o$), electric quadrupole E2 ≈ {($q a_o$) E1} and magnetic dipole M1 ≈ {($\alpha/2$) E1}, where the fine-structure constant $\alpha \approx 1/137$. In the visible and soft x-ray region, E1 events are dominant in the light-matter interaction, unless they are forbidden by symmetry and nominally weaker events are allowed to come to the fore. Evidently, pure transitions like E1-E1 and, also, E2-E2 have even parity. Only mixed terms can give odd parity. Optical activity and NCD arise from parity-odd events, including E1-E2 and E1-M1 [3].

$G$ is a scalar product of a photon tensor and an electronic tensor (multipole), which means both tensors respect the same discrete symmetries. Electronic multipoles are constructed from the operators for position **R** (time-even, parity-odd), spin **S** and orbital moment **L** (both time-odd and parity-even), and anapole moment **Ω** (time-odd, parity-odd). The latter is epitomized in the toroidal electric current [16]. When a pure transition (parity-even) is time-odd the material tensor also needs to be time-odd (magnetic polarization), which allows for the Faraday effect and MCD in any spatial symmetry group. In a mixed transition (parity-odd) the NCD reverses the rotation angle (handedness) of both the matter and the light, leaving the total system invariant. In this case, the multipole can be described by pseudo-tensors $\mathbf{U}^K$ of even rank $K$. In particular, $\mathbf{U}^0$ in the E1-M1 event is the chiral monopole, and it is multiplied by the photon helicity when it appears in the scattering amplitude. When a mixed term is time-odd (magneto-chiral dichroism) the electronic multipole is likewise time-odd, and it is satisfied by the anapole $\boldsymbol{\Omega}$.

In this paper, we extend the established concepts for ordinary dichroic signals to tertiary events of the generic form shown in Fig. 1 and (A4). On this basis, a theory of dichroism in the SHG response clearly needs a different approach than used for ordinary NCD. After a short review of this NCD signal in terms of electronic multipoles $\langle \mathbf{U}^K \rangle$ in Sec. II, we present in Sec. III our results for the MCD and NCD signals in the SHG response. Equivalent electronic operators for the signals appear in Sec. IV, followed by conclusions and a discussion in Sec. V. An Appendix contains technical details and auxiliary information about our method of working [16].

## II. ORIENTATION TO DICHROIC SIGNALS

To begin with, NCD signals result from setting a chiral structure in a handed environment created by circularly polarized light [3, 16]. Photon polarization is here described by standard Stokes parameters reviewed in the Appendix [16, 17]. Specifically, helicity $\equiv P_2$ is the average value of $\boldsymbol{\sigma} \cdot \mathbf{p}$, where $\boldsymbol{\sigma}$ and $\mathbf{p}$ are photon spin and linear momentum, respectively. The electronic structure illuminated by circularly polarized photons is defined in terms of polar spherical multipoles $\langle \mathbf{U}^K_Q \rangle$ with integer rank $K$ and projection $Q$ (with $-K \leq Q \leq +K$). Angular brackets about the tensor operator denote its expectation value in the ground-state. We use an electronic structure factor $\Psi^K_Q = \sum_d \langle \mathbf{U}^K_Q \rangle_d$, where the sum is over all sites in the sample occupied by absorbing ions. NCD signals calculated with $\Psi^K_Q$ can be different from zero for chiral structures, for they must have a form $[P_2 \Psi^K_Q]$ unchanged by spatial inversion (a specific example of Neumann's Principle [2, 18]). For a crystalline sample, environments at the different sites are related by operations that appear in the relevant space group. An individual polar (parity-odd) multipole $\langle \mathbf{U}^K_Q \rangle$ must obey site symmetry that can be more restrictive than the crystal class (point group), with matching site symmetry and crystal class an exceptional case. The dipole $\langle \mathbf{U}^1_Q \rangle$ is the electric polarization in the electronic ground-state. (Spherical and Cartesian components of a three-dimensional vector $\mathbf{R} = (x, y, z)$ are related by $x = (1/\sqrt{2}) (\mathbf{R}_{-1} - \mathbf{R}_{+1})$, $y = (i/\sqrt{2}) (\mathbf{R}_{-1} + \mathbf{R}_{+1})$, $z = \mathbf{R}_0$.)

NCD signals use multipoles with even rank. For the E1-E2 absorption event, however, the triangle rule excludes $K = 0$. The NCD signal is thus $F(1, 2) = [P_2 \Psi^2_0(\mathbf{U})]$, where the light propagates along the $z$-axis (projection $Q = 0$). For the moment, we focus on the structure of the dichroic signal. In doing so, we set aside numerical factors and radial integrals discussed in the Appendix [18-21]. A dimensionless NCD signal from the E1-M1 event is $F(1, 1) = [P_2 \{\sqrt{2} \Psi^0_0(\mathbf{U}) - \Psi^2_0(\mathbf{U})\}]$. Orbitals in the M1 absorption event possess the same angular momentum, because the magnetic moment operator is diagonal in this basis. A careful, quantum-mechanical calculation of E1-M1 absorption by gallium ferrate is reported in Ref. [22].

We conclude our orientation to NCD signals expressed by electronic multipoles with a specific example using spatial symmetry $mm2$ ($C_{2v}$), which is frequently encountered in applications. Realization as a nanoparticle is represented by an ellipsoidal-shaped dot on a substrate, with axis of symmetry normal to the substrate that breaks inversion symmetry [5, 6]. General selection rules derived from spatial symmetry are provided in the Appendix.

Consider, instead, a crystalline sample with resonant ions at sites 8$b$ in the non-centrosymmetric polar space group $Cmc2_1$ (#36, crystal axes $a \neq b \neq c$). Fig. 2 illustrates the crystal structure of $Ca_3Ru_2O_7$ with Ru in sites 8b, using the non-standard setting $Bb2_1m$, together with the motif of Ru axial dipoles and anapoles that develops spontaneously below $\approx 48$ K. Ruthenium sites have no symmetry that limits $K$ or $Q$. The electronic structure factor for the parent structure is,

$$\Psi^K{}_Q(Cmc2_1) = 2\,[1 + (-1)^Q]\,[\langle \mathcal{O}^K{}_Q \rangle + \sigma_\pi\,(-1)^K \langle \mathcal{O}^K{}_{-Q} \rangle]. \tag{1}$$

Here, $\sigma_\pi = +1$ ($-1$) for axial (polar) multipoles. Evidently, $\Psi^K{}_Q(Cmc2_1)$ is zero unless $Q$ is even.

With $\sigma_\pi = -1$ in Eq. (1), we find $\Psi^1{}_0 = 8\langle \mathbf{U}^1{}_0 \rangle$ and the material is ferroelectric with polarization parallel to the z-axis that coincides with the crystal c-axis in Fig. 2. NCD signals are zero in the nominal setting of the crystal, because $\Psi^K{}_0 = 0$ for $K$ even. Non-zero signals are available when the crystal axes ($a$, $b$, $c$) do not coincide with Cartesian axes ($x$, $y$, $z$) defined by the beam of light. Specifically, let the $c$-axis and $x$-axis coincide ($c$-axis parallel to σ-polarization, cf. Appendix), and rotate the crystal by an angle ψ about the $c$-axis. One finds $\Psi^0{}_0(\mathbf{U}) = 0$. On the other hand, $\Psi^2{}_0(\mathbf{U}) \propto \sin(2\psi)\,\langle \mathbf{U}^2{}_{+2} \rangle''$, where the $a$-axis is normal to the $x$-$y$ plane for $\psi = 0$, and the NCD can be different from zero if the imaginary part of the quadrupole $\langle \mathbf{U}^2{}_{+2} \rangle$ is different from zero. The two-fold rotation symmetry in ψ is a consequence of a diad axis of rotation symmetry in the space group.

Magnetic circular dichroism can likewise be described in terms of electronic multipoles [13-16]. In the next section, we demonstrate that both NCD and MCD of the SHG response can be so described.

### III. CIRCULAR DICHROISM OF THE SHG RESPONSE

Dichroic signals relate to the amplitude for photon scattering $G_{\mu\nu}$ derived by perturbation theory from QED and salient steps are gathered in the Appendix [17, 19, 23, 24]. In the present work, $G_{\mu\nu}$ is truncated at the level of E1, M1, and E2 absorption events. They are labelled by polarizations $\mu$ and $\nu$ of two photons with energy E and one photon with polarization $\varepsilon'$ and energy 2E, which are depicted in Fig. 1. A dichroic signal $F$ is the dimensionless form of $G_{\mu\nu}$ following the average with respect to $\mu$ and $\nu$ that renders it a function of Stokes parameters.

A consequence of the truncation is that MCD [25] and NCD are allowed in our investigation, and no other signals are present. This finding can be deduced by symmetry considerations alone, without specific calculations, as we now demonstrate. From (A2), E1 is independent of the photon wavevector, while M1 and E2 are both proportional to **q**. The parity-even absorption event E1′-E2-E1 creates an MCD of the form [**q** $P_2$ ⟨$\mathbf{T}^K$⟩], where ⟨$\mathbf{T}^K$⟩ is a parity-even multipole, cf. (3). Let us test the expression for MCD, a scalar observable, against requirements imposed by discrete symmetries. We will discover restrictions on the rank and projections of the multipole. The dichroic signal is not changed by spatial inversion or time-reversal, for example, in keeping with Neumann's Principle [18]. The product (**q** $P_2$) is unchanged by spatial inversion, since **q** is polar and $P_2$ is a pseudo-scalar that is likewise odd with respect to inversion. The Stokes parameter $P_2$ is time-even whereas **q** is time-odd. In consequence, the signal in question is unchanged by time-reversal given that ⟨$\mathbf{T}^K$⟩ is time-odd (magnetic). In addition, ⟨$\mathbf{T}^K$⟩ must be odd with respect to twofold rotation about an axis normal to **q**, a rotation denoted $2_\perp$, since $P_2$ is unchanged by $2_\perp$. (Note that $2_\perp$ is equivalent to time-reversal.) The Stokes parameter is similarly unchanged by a twofold rotation around **q**, and ⟨$\mathbf{T}^K$⟩ must possess the same rotational symmetry. Looking ahead, our result in Eq. (3) is $F$(MCD) ∝ [$q_0$ $P_2$ ⟨$\mathbf{T}^K_0$⟩] with $K$ odd and $Q = 0$, fulfils all symmetry that has been mentioned. Turning to parity-odd events, E1′-E1-E1 is responsible for an NCD signal [$P_2$ ⟨$\mathbf{U}^K$⟩]. The aforementioned invariance of $P_2$ with respect to $2_\perp$ means that components of the multipole allowed in $F$(NCD) are ⟨$\mathbf{U}^K_0$⟩ with $K$ even.

Symmetry-based arguments of the type just encountered show that no other dichroic signals are allowed with the truncated photon amplitude. By way of a signal actually excluded by symmetry, consider magneto-chiral dichroism with a generic symmetry [**q** ⟨$\mathcal{O}^K$⟩] and ⟨$\mathcal{O}^K$⟩ both parity-odd and time-odd (magneto-electric). No such symmetry can be constructed from E1, M1, and E2.

In subsequent calculations, the primary beam is parallel to the z-axis. Primary polarization vectors are **μ** = (1, 0, 0) or **ν** = (0, 1, 0). The secondary beam is inclined to the *z*-axis and polarization vectors **ε**′ include a component (0, 0, 1). A sketch is shown in Fig. 3.

Photon variables are assembled in a spherical tensor, i.e., a function that behaves like a spherical harmonic. Since $G_{\mu\nu}$ is a scalar quantity the electronic multipole and photon tensor, $\mathbf{C}^K_Q(\mu\nu)$ say, must create a scalar product $G_{\mu\nu} \propto \{(-1)^Q \mathbf{C}^K_Q(\mu\nu) ⟨\mathcal{O}^K_{-Q}⟩\}$, with sums on repeated labels. Our task is to calculate the electronic multipole and photon tensor for tertiary events in SHG. We start by considering the parity-even E1′-E2-E1 for $F$(MCD), where $\mathbf{C}^K_Q(\mu\nu)$ and ⟨$\mathcal{O}^K_Q$⟩ are explicit in (2) as $\mathbf{B}^K_Q(p; \mu\nu)$ and ⟨$\mathbf{T}^K_Q(p)$⟩, respectively. Likewise, $\mathbf{A}^K_Q(p; \mu\nu)$ and ⟨$\mathbf{U}^K_Q(p)$⟩ in (4) for parity-odd E1′-E1-E1.

Our calculations show that discrete symmetries are not unique labels of tertiary multipoles. This contrasts with ordinary absorption using dyadic multipoles (E1-E1 and E2-E2), where even-rank multipoles are time-even (charge-like) and odd-rank are time-odd (magnetic). This particular rule no longer holds for parity-odd binary events, however. In this case, there are simply two types of dyadic multipoles distinguished by their time signatures. For core excitations, parity-even dyadic multipoles display celebrated sum-rules that enable spin and orbital moments to be extracted from integrated absorption profiles [13-15, 26]. Equivalent sum-rules in dyadic parity-odd events are more complicated and less useful [20, 21].

The amplitude for photon scattering derived from parity-even E1′-E2-E1 is,

$$G_{\mu\nu} = \sum_{K,Q,p} i^{K+1} (-1)^Q \mathbf{B}^K{}_Q(p; \mu\nu) \langle \mathbf{T}^K{}_{-Q}(p) \rangle, \qquad (2)$$

$$\mathbf{B}^K{}_Q(p; \mu\nu) = \sum_{q,\gamma} \varepsilon'_\gamma \mathbf{N}^p{}_q(\mu\nu) \ (1\gamma\, pq\,|\, KQ),$$

with $\{\mathbf{B}^K{}_Q(p; \mu\nu)\}^* = (-1)^{K+Q} \mathbf{B}^K{}_{-Q}(p; \mu\nu)$,

$$\mathbf{N}^p{}_q(\mu\nu) = \sum_{\alpha,\beta} \mu_\alpha \nu_\beta \ (2\alpha\, 1\beta\,|\, pq),$$

and $p = 1, 2, 3$ labels modes available in the primary beam. Clebsch-Gordan coefficients ($a\alpha\, b\beta\,|\, KQ$) in Eq. (2) are purely real [27, 28, 29]. Matrix elements of $\mathbf{T}^K{}_{-Q}(p)$ are the subject of the next section. The quadrupole operator for E2 is a spherical harmonic of rank 2. The MCD dichroic signal $\propto \{iP_2 [G_{\mu\nu} - G_{\nu\mu}]\}$ [16] is found to exist for $p = 2$ and,

$$F(\text{MCD}) = q_0\, P_2\, \{\sqrt{2}\langle \mathbf{T}^1{}_0(2) \rangle + \sqrt{3}\langle \mathbf{T}^3{}_0(2) \rangle\}, \qquad (3)$$

where $q_0 \equiv q_z$ represents the primary wavevector. Our Hermitian multipoles with projection $Q = 0$ are purely real.

The amplitude for photon scattering derived from E1′-E1-E1 is slightly different from Eq. (2), namely,

$$G_{\mu\nu} = \sum_{K,Q,p} i^{K+1} (-1)^Q \mathbf{A}^K{}_Q(p; \mu\nu) \langle \mathbf{U}^K{}_{-Q}(p) \rangle, \qquad (4)$$

$$\mathbf{A}^K{}_Q(p; \mu\nu) = i \sum_{q,\gamma} \varepsilon'_\gamma \mathbf{X}^p{}_q(\mu\nu) \ (1\gamma\, pq\,|\, KQ),$$

with $\{\mathbf{A}^K{}_Q(p; \mu\nu)\}^* = (-1)^{K+Q} \mathbf{A}^K{}_{-Q}(p; \mu\nu)$,

$$\mathbf{X}^p{}_q(\mu\nu) = \sum_{\alpha,\beta} \mu_\alpha \nu_\beta \, (1\alpha\, 1\beta\,|\,pq).$$

Not surprisingly, the NCD signal $\{iP_2\,[G_{\mu\nu} - G_{\nu\mu}]\}$ has the structure derived for E1-E2 and E1-M1 absorption events. It exists for $p = 1$ and,

$$F(\text{NCD}) = P_2 \langle \mathbf{U}^2{}_0(1) \rangle. \tag{5}$$

Notably, $\mathbf{U}^2{}_0(1)$ is found to be a dyadic operator of electronic variables $\mathbf{L}$ and $\mathbf{R}$. This surprising result is a consequence of the commutation relation $(\mathbf{L} \times \mathbf{L}) = i\mathbf{L}$, meaning that $\mathbf{U}^2{}_0(1)$ is purely quantum mechanical with no classical analogue.

## IV. CALCULATIONS

Derivations of Eqs. (2) and (4) follow the treatment of optical transition probabilities by Judd and Ofelt [30-33]. It follows from (A3) that $G_{\mu\nu}$ is a product of three matrix elements. Each one is of the form $\langle JM|O|J'M'\rangle$, where $O$ can be E1, E2, or M1, and $J$ is the total angular momentum of an atomic state and $M$ a magnetic projection ($J$ and $M$ are half-integers), i.e., $G_{\mu\nu} \propto \{\langle JM|O|jm\rangle\langle jm|P|j'm'\rangle\langle j'm'|Q|J'M'\rangle\}$. Judd and Ofelt investigated the result of integrating out intermediate degrees of freedom, and this we have accomplished for $G_{\mu\nu}$. Fundamental results that flow from the algebra are (i) tertiary matrix elements create a spherical tensor $\mathcal{O}^K{}_Q$, and (ii) $G_{\mu\nu}$ takes the form $G_{\mu\nu} \propto \{(-1)^Q \, \mathbf{C}^K{}_Q(\mu\nu) \, \langle \mathcal{O}^K{}_{-Q}\rangle\}$, where the tensor $\mathbf{C}^K{}_Q(\mu\nu)$ represents all relevant photon variables, and there are sums on repeated labels. Creation of the spherical tensor, result (i), has profound implications, because the full weight of Racah algebra can be deployed thereafter. Specifically, matrix elements of $\mathcal{O}^K{}_Q$ satisfy the Wigner-Eckart theorem [27, 33, 34, 35]. Moreover, rotation operations have simple results, e.g., $2_\perp \mathcal{O}^K{}_Q = (-1)^K \mathcal{O}^K{}_{-Q}\,[(-1)^{K+Q} \mathcal{O}^K{}_{-Q}]$ for the $x$-axis [$y$-axis]. Result (i) is achieved by sums on intermediate projections $m$ and $m'$ alone, and a loss of geometric properties of the tertiary matrix element is what it costs, i.e., angular anisotropy of intermediate states is discarded. In the following example, additional sums on $j$ and $j'$ lead to equivalent operators with appealing properties [36].

The Wigner-Eckart theorem is,

$$\langle JM|\mathcal{O}^K{}_Q|J'M'\rangle = (-1)^{J-M} \begin{pmatrix} J & K & J' \\ -M & -Q & M' \end{pmatrix} (J\|\mathcal{O}^K\|J'). \tag{6}$$

A reduced matrix-element $(J\|\mathcal{O}^K\|J')$, which is also called a double-barred matrix element, obeys two identities. First, the reduced matrix-element (RME) of a Hermitian operator obeys [29, 35, 37],

$$(\sigma lJ\|\mathcal{O}^K\|\sigma l'J') = (-1)^{J-J'} (\sigma l'J'\|\mathcal{O}^K\|\sigma lJ)^*, \tag{7}$$

with σ = 1/2. Henceforth, we simplify notation by omitting the spin label σ that is diagonal in all RMEs for coupled states with quantum labels σ, $l$, and $J = (l \pm \sigma)$. Furthermore [29, 35],

$$(lJ||\mathcal{O}^K||l'J') = (-1)^K \sigma_\pi \sigma_\theta (lJ||\mathcal{O}^K||l'J')^*, \tag{8}$$

where $\sigma_\theta = \pm 1$ is the time signature of $\mathcal{O}^K$. We pause to note that in J-J coupling the RME of a spherical harmonic $\mathbf{C}^k(\mathbf{R})$ depends only on the $J$s and not on the $\sigma l$ of the electrons. (In consequence, coefficients of the Slater integrals $F^k$ for the interaction between two $p_{3/2}$ electrons are the same as between two $d_{3/2}$ or between a $p_{3/2}$ and a $d_{3/2}$ [37].)

Let $Z^K(lJ, l'J')_1$ be a purely real RME and $\sigma_\pi \sigma_\theta = -1$, which are the correct signatures for both $\mathbf{T}^K$ and $\mathbf{U}^K$. The second of our two identities is satisfied by $(lJ||\mathcal{O}^K||l'J') = i^{K+1} Z^K(lJ, l'J')_1$. To comply with the first identity,

$$(lJ||\mathbf{U}^K(p)||l'J') = i^{K+1} [Z^K(lJ, l'J')_1 - Z^K(lJ, l'J')_2], \tag{9}$$

for NCD ($K$ even) and,

$$(lJ||\mathbf{T}^K(p)||l'J') = i^{K+1} [Z^K(lJ, l'J')_1 + Z^K(lJ, l'J')_2], \tag{10}$$

for MCD ($K$ odd). In both cases, $K$ even and $K$ odd, there is a conjugate RME defined by $Z^K(l'J', lJ)_1 = (-1)^{J-J'} Z^K(lJ, l'J')_2$.

A result for $Z^K(lJ, l'J')_1$ is obtained from $\{\langle JM|E1|jm\rangle\langle jm|Ek|j'm'\rangle\langle j'm'|E1|J'M'\rangle\}$ after completing sums on projections $m$ and $m'$, which appear only in 3$j$-symbols through three applications of the Wigner-Eckart theorem. Specifically,

$$Z^K(lJ, l'J')_1 = (-1)^{J-j+p+k} (lJ ||\mathbf{R}||l''j) (l''j||\mathbf{C}^k||l'''j') (l'''j'||\mathbf{R}||l'J')$$

$$\times \begin{Bmatrix} 1 & p & K \\ J' & J & j \end{Bmatrix} \begin{Bmatrix} k & 1 & p \\ J' & j & j' \end{Bmatrix}, \tag{11}$$

with $\mathbf{C}^1(\mathbf{R}) = \mathbf{R}$. The energy levels for Eq. (11) are depicted in Fig. 1. Since the spherical harmonic $\mathbf{C}^k$ is Hermitian and the RME is purely real it follows that $(lj||\mathbf{C}^k||l'j') = (-1)^{j-j'} (l'j'||\mathbf{C}^k||lj)$, and the derivation of $Z^K(lJ, l'J')_2$ from Eq. (11) is then straightforward. In order to perform sums on $j$ and $j'$ in Eq. (11) we need an explicit result for $(lj||\mathbf{C}^k||l'j')$. In our chosen coupling scheme, often labelled s-l coupling (as opposed to l-s coupling [27, 29, 35]),

$$(\sigma l j\|\mathbf{C}^k\|\sigma l'j') = (-1)^{\sigma + l' + j + k} \begin{Bmatrix} l & j & \sigma \\ j' & l' & k \end{Bmatrix} (l\|\mathbf{C}^k\|l'). \tag{12}$$

We go on to find,

$$z^K(lJ, l'J')_1 = \sum_{j,j'} Z^K(lJ, l'J')_1 = (-1)^{\sigma + J + p + k + K + l + l' + l''} [(2J + 1)(2J' + 1)]^{1/2}$$

$$\times (l\|\mathbf{R}\|l'') (l''\|\mathbf{C}^k\|l''') (l'''\|\mathbf{R}\|l') \begin{Bmatrix} l & J & \sigma \\ J' & l' & K \end{Bmatrix} \begin{Bmatrix} K & l & l' \\ l'' & p & 1 \end{Bmatrix} \begin{Bmatrix} p & l'' & l' \\ l''' & 1 & k \end{Bmatrix}. \tag{13}$$

It remains to write this key result in a perspicuous form, namely, equivalent operators.

For the NCD signal we evaluate Eq. (13) with $k = 1$, $p = 1$ (E1′-E1-E1), and define an RME, $z^K(l, l')_1$, consistent with the coupling scheme explicit in Eq. (12),

$$z^K(lJ, l'J')_1 = (-1)^{\sigma + l' + J + K} \begin{Bmatrix} l & J & \sigma \\ J' & l' & K \end{Bmatrix} z^K(l, l')_1, \tag{14}$$

and here $K = 2$. Likewise, for the definition of $z^2(l, l')_2$. On inserting the two results in Eq. (9),

$$(\sigma lJ\|\mathbf{U}^2(1)\|\sigma l'J') = (-1)^{\sigma + l' + J} \begin{Bmatrix} l & J & \sigma \\ J' & l' & 2 \end{Bmatrix} (l\|\mathbf{U}^2(1)\|l'), \tag{15}$$

with,

$$(l\|\mathbf{U}^2(1)\|l') = i(1/\sqrt{2}) [(2l + 1)(2l' + 1)]^{-1}$$

$$\times [(2l + 1)(l\|\{\mathbf{R} \otimes \mathbf{L}\}^2\|l') - (2l' + 1)(l\|\{\mathbf{L} \otimes \mathbf{R}\}^2\|l')]. \tag{16}$$

In Eq. (16) we introduce a tensor product to define an equivalent operator using the definition [28],

$$\{\mathbf{A}^a \otimes \mathbf{B}^b\}^K_Q = \sum_{\alpha,\beta} \mathbf{A}^a_\alpha \mathbf{B}^b_\beta (a\alpha\, b\beta \mid KQ). \tag{17}$$

Clebsch-Gordan coefficients in Eqs. (2), (4), and (17) are purely real, and related to the 3$j$ symbol that occurs in the Wigner-Eckart theorem of Eq. (6) [27, 28, 29],

$$(a\alpha\, b\beta \mid KQ) = (-1)^{-a + b - Q} \sqrt{(2K + 1)} \begin{pmatrix} a & b & K \\ \alpha & \beta & -Q \end{pmatrix}. \tag{18}$$

Results in Eqs. (15) and (16) completely define the multipole $\langle \mathbf{U}^2(1) \rangle$ for the NCD signal derived from the SHG response using E1′-E1-E1.

A calculation for E1′-M1-M1, similar to the foregoing one, shows that the event does not produce NCD, i.e., the corresponding $(l\|\mathbf{U}^2(1)\|l') = 0$. RMEs for spin and orbital operators in M1 $= (\mathbf{L} + 2\mathbf{S})$ obey different relations [27, 29, 35]. While the RME of $\mathbf{L}$ satisfies Eq. (12) the same is not true for $\mathbf{S}$, and the difference adds great complexity to the calculation of $(l\|\mathbf{U}^2(1)\|l')$, which we do not report.

Turning to MCD, we evaluate Eq. (13) with $k = 2$, $p = 2$ and insert results in

$$(l\|\mathbf{T}^K(2)\|l') = [z^K(l, l')_1 + (-1)^{l - l'} z^K(l', l)_1], \tag{19}$$

with $K = 1$ and 3, and $z^K(l, l')_1$ is defined in accord with Eq. (14). More progress is achieved with an orbital anapole [35],

$$\mathbf{\Omega} = (\mathbf{L} \times \mathbf{R}) - (\mathbf{R} \times \mathbf{L}) = i[\mathbf{L}^2, \mathbf{R}], \tag{20}$$

and the RME,

$$(l\|\mathbf{\Omega}\|l') = i(l\|\mathbf{R}\|l')\,[l(l + 1) - l'(l' + 1)], \tag{21}$$

is purely imaginary and symmetric with respect to an interchange of $l$ and $l'$, which are the opposite properties of $(l\|\mathbf{R}\|l')$. An anapole, which appears in the magnetic motif depicted in Fig. 2, is magnetic ($\sigma_\theta = -1$) and polar ($\sigma_\pi = -1$), and a product of $\mathbf{L}$ and $\mathbf{\Omega}$ is time-even and polar. Returning to Eq. (13), we use identities,

$$(l''\|\mathbf{C}^2\|l''')\,(l'''\|\mathbf{R}\|l')\begin{Bmatrix} 2 & l'' & l' \\ l''' & 1 & 2 \end{Bmatrix} \tag{22}$$

$$= \pm(i/(2\sqrt{5}))\,[(l' + 1)(2l' + 1)]^{-1}\,(l''\|\{\mathbf{L} \otimes \mathbf{\Omega}\}^2\|l'),\ \ l'' = l' + 1,\ l''' = l' \pm 1$$

$$= \pm(i/(2\sqrt{5}))\,[l'(2l' + 1)]^{-1}\,(l''\|\{\mathbf{L} \otimes \mathbf{\Omega}\}^2\|l'),\ \ l'' = l' - 1,\ l''' = l' \mp 1.$$

The identities provide a correspondence,

$$z^K(l, l')_1 \propto (l\|\mathbf{R}\|l'')\,(l''\|\{\mathbf{L} \otimes \mathbf{\Omega}\}^2\|l')\begin{Bmatrix} K & l & l' \\ l'' & 2 & 1 \end{Bmatrix}. \tag{23}$$

Note that $l = l'$ for $K = 1$. Interestingly, an operator equivalent for $z^K(l, l')_1$ is achievable, in an essentially semi-classical limit, when numerical values of $l'$ and $(l' + 1)$ in the pre-factors to the two identities are regarded equal. In this extreme limit, $z^K(l, l')_1 \propto (l\|\{\mathbf{R} \otimes \{\mathbf{L} \otimes \mathbf{\Omega}\}^2\}^K\|l')$. Moreover, a sum on $l''$ in Eq. (23) yields precisely the same result [27], while no such reduction is required to achieve Eq. (16).

## V. CONCLUSIONS AND DISCUSSION

In summary, we presented a theory for circular dichroism in the second-harmonic generation (SHG) response. Using theoretical techniques from atomic physics, explicit expressions for electronic multipoles in natural circular (NCD) and magnetic circular (MCD) dichroism are derived. SHG is a tertiary photon event which goes beyond the parity- and time-reversal signature of dyadic photon events encountered in the well-known Kramers-Heisenberg dispersion formula.

In the more general case, each term in a standard expansion of the light-matter scattering amplitude is a scalar product of a photon tensor and an electronic tensor (multipole). In consequence, both tensors respect the same discrete symmetries to respect Neumann's Principle [2]. Our truncated light-matter amplitude based on (A2) includes, the electric dipole E1, electric quadrupole E2, and magnetic dipole M1, and the latter two events scale with the primary photon wavevector $\mathbf{q}$. Circular polarization in the photon tensor is evoked by the pseudo-scalar (parity-odd) Stokes parameter $P_2$, which is time-even.

### A. Expressions for dichroic signals

A parity-odd tertiary event E1′-E1-E1 used by the SHG response is responsible for NCD with electronic symmetry $\{P_2 \langle \mathbf{U}^K \rangle\}$, where $\mathbf{U}^K$ is a polar (parity-odd) and time-even (non-magnetic) multipole with even rank $K$ [see Eq. (5)]. Equivalent electronic operators for multipole operators are constructed from position $\mathbf{R}$ (time-even, parity-odd), spin and orbital dipoles $\mathbf{S}$ and $\mathbf{L}$ (time-odd, parity-even), and an anapole moment $\mathbf{\Omega}$ (time-odd, parity-odd). The equivalent operator is a simple quadrupole (tensor product) $[i\{\mathbf{R} \otimes \mathbf{L}\}^2]$ for NCD derived from E1′-E1-E1. No NCD is found to arise from the analogous process using magnetic dipole events in the primary absorption, namely, E1′-M1-M1.

Turning to a tertiary parity-even event, E1′-E2-E1 in SHG yields a MCD signal with generic symmetry $\{\mathbf{q} P_2 \langle \mathbf{T}^K \rangle\}$. The product $(\mathbf{q} P_2)$ is time-odd and unchanged by spatial inversion, matching discrete symmetries of magnetic multipoles $\langle \mathbf{T}^K \rangle$ [see Eq. (3)]. Equivalent operators for $\mathbf{T}^K$ are tensor products $\{\mathbf{R} \otimes \{\mathbf{L} \otimes \mathbf{\Omega}\}^2\}^K$ with $K$ odd.

### B. Digression

We can position our calculation of equivalent operators for the SHG response with derivations of sum rules for ordinary dichroic signals that use dyadic matrix elements. In the original formulation, sum rules are a consequence of a closure theorem for complete sets of atomic states [13-15, 26, 38, 39]. A dyadic matrix element is thereby transformed to standard reduced matrix elements (RMEs) usually denoted $\mathbf{W}^{(a,b)K}(lJ, l'J')$, for valence states with quantum numbers *lJ*, *l'J'*, with a triple sum on labels *a* (spin), *b* (angular momentum), and *K* (tensor rank). Coefficients in the sums possess quantum labels for the core state. Analysis of the factorization, of valence and core state degrees of freedom, yields sum rules for integrated intensities of an energy profile. The closure theorem is equivalent to a sum over projections of intermediate states, denoted in Sec. IV by *m*, while *j* labels the core state, e.g., $L_2$, $M_3$, etc. [16]. Evidently, a further sum over total angular momenta of intermediate states, *j*, removes all core-state degrees of freedom. Integrating out intermediate degrees of freedom, as we have described, is the central achievement of Judd and Ofelt in their celebrated work on optical transition probabilities [30, 31]. At the opposite extreme, when all core-state degrees of freedom *j*, *m* are retained, a dyadic matrix element can be represented by spherical tensors [40], and the same type of representation holds for parity-odd dyadic matrix elements [41].

Spin degrees of freedom are not engaged in electronic events E1 and E2 explicitly examined in our communication. In consequence, the atomic RME $\mathbf{W}^{(a,b)K}(lJ, l'J')$ occurs here with $a = 0$, i.e., effective operators use $\mathbf{W}^{(0,K)K}(lJ, l'J')$. The full RME, with $a = 1$ and *b* different from *K*, occurs in amplitudes that use the magnetic dipole. Our calculations have shown that the SHG response derived from E1′-M1-M1 does not provide NCD, although we do not report details of the attendant algebra in light of its complexity.

## APPENDIX

Ordinary dichroic signals use dyadic matrix elements that arise in the Kramers-Heisenberg amplitude [12]. Let us label virtual intermediate states involved in a photon absorption event by a composite quantum number λ; it is shorthand for atomic numbers σ, *l*, *j*, *m*, in Section IV. Unlike the initial and final states of the sample, intermediate states are not from the equilibrium configuration (ground state) and they decay on a timescale ≈ ℏ/Γ where Γ is the total width of the resonance. Photon variables, **ε** and **q** (**ε**′ and **q**′) denote the primary (secondary) polarization vector and wavevector, respectively, while the magnitude of the wavevector q = (E/ ℏ c). The scattering length in the vicinity of a resonance is derived using QED [17, 19],

$$f \approx - (r_e/m) \sum_\lambda \langle \{\boldsymbol{\varepsilon}' \cdot \mathfrak{J}(-\mathbf{q}')|\lambda\rangle \langle\lambda| \boldsymbol{\varepsilon} \cdot \mathfrak{J}(\mathbf{q})\}\rangle/[E - \Delta + i\Gamma_\lambda/2], \quad (A1)$$

where $r_e \approx 0.282 \times 10^{-12}$ cm, m is the mass of an electron, and the sum of intermediate states is limited to those that contribute at the resonance Δ, and the primary photon energy E ~ Δ. Angular brackets denote a time average, or expectation value, of the enclosed operators. A matrix element of the current operator between an intermediate state and a component of the ground state |g⟩ is,

$$\langle\lambda| \boldsymbol{\varepsilon} \cdot \mathfrak{J}(\mathbf{q})|g\rangle \approx (im \Delta/ \hbar) \sum_j \langle\lambda| \boldsymbol{\varepsilon} \cdot \mathbf{R}_j [1 + (i/2) \mathbf{q} \cdot \mathbf{R}_j] |g\rangle + (i\hbar/2) \sum_j \langle\lambda| \mathbf{q} \times \boldsymbol{\varepsilon} \cdot [\mathbf{L} + 2\mathbf{S}]_j |g\rangle, \quad (A2)$$

where the sum is over all electrons participating in the absorption event, and $\Delta = E_\lambda - E_g \geq 0$. Electronic operators in (A2) are dipole (**R**), spin (**S**) and orbital angular momentum (**L**). From left to right, E1 $\propto \boldsymbol{\varepsilon} \cdot \mathbf{R}$, E2 $\propto (\boldsymbol{\varepsilon} \cdot \mathbf{R} \mathbf{q} \cdot \mathbf{R})$ and M1 $\propto (\mathbf{q} \times \boldsymbol{\varepsilon} \cdot [\mathbf{L} + 2\mathbf{S}])$. A contribution to (A1) depends on primary and secondary polarization vectors, $\boldsymbol{\varepsilon}$ and $\boldsymbol{\varepsilon}'$, and together with **q**, limited to E2 and M1, they form a spherical tensor $\mathbf{C}^K{}_Q(\mu\nu)$, where $\mu$, $\nu$ label polarization states. After integrating out intermediate degrees of freedom [30, 31], $f \approx - (\Delta\, r_e\, F_{\mu\nu})\,/[E - \Delta + i\Gamma/2]$ where $F_{\mu\nu}$ is the dimensionless form of $G_{\mu\nu} \propto \{(-1)^Q\, \mathbf{C}^K{}_Q(\mu\nu)\, \langle \mathcal{O}^K{}_{-Q}\rangle\}$ with sums on repeated labels. We define a dichroic signal to be the appropriate part of $F_{\mu\nu}$ after it is averaged over polarization vectors in the primary photon beam. The precise result of the ordinary NCD dichroic signal is quoted in §II. An atomic multipole $\langle \mathcal{O}^K{}_Q\rangle$ is derived from electronic operators **R**, **L** and **S**, and it possesses discrete symmetries with respect to space and time. Spherical harmonics and electronic multipoles are Hermitian with $\langle \mathcal{O}^K{}_Q\rangle^* = (-1)^Q \langle \mathcal{O}^K{}_{-Q}\rangle$, and $\langle \mathcal{O}^K{}_Q\rangle = [\langle \mathcal{O}^K{}_Q\rangle' + i\langle \mathcal{O}^K{}_Q\rangle'']$ defines real and imaginary parts. A spherical harmonic is used for the angular dependence of the dipole E1 in our dimensionless $F_{\mu\nu}$, while the magnitude of the dipole appears in a radial integral, about which we say more below. Likewise, for a product of dipole operators in E2. Multipoles $\langle \mathbf{U}^K{}_Q\rangle$ for NCD are parity-odd (polar) and time-even (non-magnetic), and $\langle \mathbf{T}^K{}_Q\rangle$ in MCD are parity-even (axial) and time-odd (magnetic).

An average over polarization vectors that reside in $\mathbf{C}^K{}_Q(\mu\nu)$ renders $f$ a function of Stokes parameters that are reviewed later in the Appendix [16, 17]. A dichroic signal is an integral of Im.($f$), taken with respect to the photon energy E, and specific to the response to polarization in the primary beam, i.e., a Stokes parameter. Attenuation (dichroism) and retardation (birefringence) in a foil are determined by the polarization dependence of its refractive index, denoted n = (n' + in''). The relation between n and $f$ is taken to be n = $(1 + (2\pi\rho/q^2)f)$ in which $\rho$ is the density of resonant ions present in the foil. According to the optical theorem, n'' = $(2\pi\,\rho/q^2)$ Im.($f$) = ($\gamma/2q$), where the attenuation coefficient, $\gamma$, has the dimension of length$^{-1}$. One readily establishes the integral relation $\int(\gamma/E)dE = (4\pi^2\, r_e\, \rho/q)\, F_{\mu\nu}$.

Inspection of (A2) shows that relative contributions to $f$ depend also on q. To create a meaningful measure of relative strengths of interference contributions it is useful to consider a dimensionless quantity $\mathfrak{R}$ that includes q and atomic radial integrals. In the case of the E1-M1 event, there is the familiar dipole radial integral from the E1-event, namely, $(\Theta|R|\Xi)$, where $\Theta$ is a valence state that carries orbital angular momentum $l$, and $\Xi$ is the intermediate state which accepts the photon and it carries orbital angular momentum $l''$, say; the two angular momenta differ by unity, as required for a parity-odd event. To the extent that radial wavefunctions in a highly ionized ion are hydrogenic in form, $(\Theta|R|\Xi)$ is proportional to $1/Z_c$ where $Z_c$ is the effective core charge seen by the jumping electron. And an explicit calculation yields $\langle 3d|R|4p\rangle = 1.3\ (a_o/Z_c)$, for example, where $a_o$ is the Bohr radius. The second radial integral $(\Theta'|\Xi)$ is the radial part of the matrix element of the magnetic moment. The valence state $\Theta'$ and $\Xi$ possess identical orbital angular momenta, because the magnetic moment operator does not change angular momentum [22]. With these definitions, $\Re(E1\text{-}M1) = [q\ (\Theta|R|\Xi)(\Theta'|\Xi)]$. Likewise, for an E1-E2 event, $\Re(E1\text{-}E2) = (\Upsilon\ [q\ (\Theta|R|\Xi)(\Theta'|R|\Xi)/a_o^2])$, where $\Theta'$ carries angular momentum $l''$ and $l + l''$ is odd. The dimensionless factor $\Upsilon = (m\ \Delta\ a_o^2/\hbar^2)$ is used to maintain relative magnitudes between magnetic and electric contributions. For parity-even events, $\Re(E1\text{-}E1) = (\Upsilon\ [(\Theta|R|\Xi)/a_o]^2)$ and $\Re(E2\text{-}E2) = (\Upsilon\ [q\ (\Theta|R^2|\Xi)/a_o]^2)$.

Prior to moving on with dichroic signals from the SHG response, we briefly review properties of the Stokes parameters [16, 17]. $P_1$ and $P_3$ are measures of the linear polarization. In the right-handed and orthogonal set of coordinates ($\boldsymbol{\sigma}, \boldsymbol{\pi}, \mathbf{q}/q$), $P_1$ describes polarization along directions at angles $\pm 45°$ to the σ-axis. The parameter $P_3$ describes polarization along the σ- and π-axis; $P_3 = +1$ corresponds to complete polarization in the σ-direction, and $P_3 = -1$ corresponds to complete polarization in the π-direction. The parameter $P_2$ measures the degree of circular polarization. Here, it is defined to be the mean value of the helicity operator. All parameters are purely real and time-even. $P_1$ and $P_3$ are true scalars, while $P_2$ is a pseudo-scalar. The parameters satisfy $(P_1^2 + P_2^2 + P_3^2) \leq 1$ and the equality is achieved for a completely polarized beam.

Third-order perturbation theory accounts for the SHG response [4]. One of three terms in the scattering length is,

$$\sum_{\lambda,\lambda'} \langle \{\boldsymbol{\varepsilon}' \cdot \mathfrak{J}(-\mathbf{q}')|\lambda\rangle \langle \lambda| \boldsymbol{\varepsilon} \cdot \mathfrak{J}(\mathbf{q})|\lambda'\rangle \langle \lambda'| \boldsymbol{\varepsilon} \cdot \mathfrak{J}(\mathbf{q})\}\rangle/\{[2E - \Delta_\lambda + i\Gamma_\lambda/2][E - \Delta_{\lambda'} + i\Gamma_{\lambda'}/2]\}, \quad (A3)$$

with $\Delta_\lambda = (E_\lambda - E_g)$. The generic form of the NCD signal from the E1'-E1-E1 response is,

$$P_2 \sum_{\lambda,\lambda'} \{\langle g|x|\lambda\rangle\langle\lambda|y|\lambda'\rangle\langle\lambda'|z|g\rangle - \langle g|y|\lambda\rangle\langle\lambda|x|\lambda'\rangle\langle\lambda'|z|g\rangle\}. \quad (A4)$$

Here, coordinates are defined by a primary beam parallel to the z-axis and σ-polarization parallel to x-axis, and a unit vector for the electronic dipole is denoted (x, y, z). The secondary wavevector, for E1', is inclined to the z-axis and its polarization vector casts a shadow on the axis. Intermediate degrees of freedom $\lambda, \lambda'$ are integrated out in the construction of $G_{\mu\nu}$ given in (2) and (4), together with photon tensors $\mathbf{B}^K_Q(p; \mu\nu)$ and $\mathbf{A}^K_Q(p; \mu\nu)$. Corresponding atomic multipoles $\langle \mathcal{O}^K_Q \rangle$ are formed with three electronic operators as one sees explicitly in (11), for example.

The geometric factor for NCD involves $K = 0^-$ and $2^-$ in $O_3$ symmetry (the superscript minus sign denotes the parity label). $K = 0^-$ ($O_3$) never branches to the total symmetric representation in any group containing inversion or reflection symmetry. Thus, breaking the inversion symmetry is conditional to observe NCD. On the other hand, $K = 2^-$ branches to a total symmetric representation in the low symmetry groups $C_{2v}$, $C_s$, $D_{2d}$ and $S_4$, so that NCD is permitted in these spatial symmetries.

**Acknowledgements**. We are grateful to Prof R. Ramesh for stimulating the reported calculations. Dr D. D. Khalyavin scrutinized our use of crystallography and prepared Fig. 2.

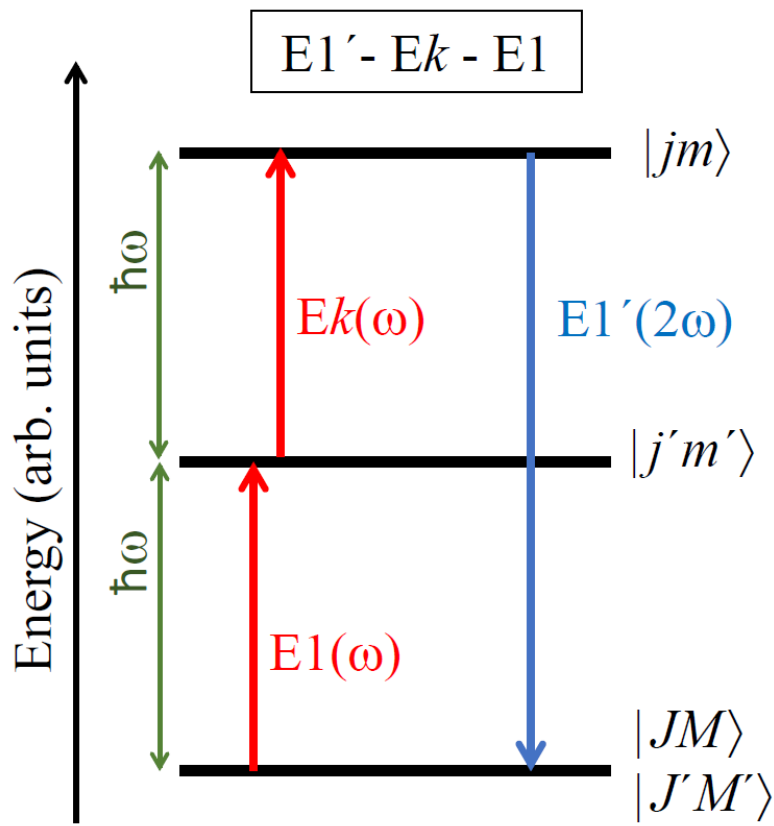

**FIG. 1.** Transitions that occur in second harmonic generation (frequency doubling) using three electronic operators E1′-E$k$-E1, where E$k$ = E1 or E2 and the primed operator E1′ relates to the secondary transition. Primary energy E = ℏω. Other events such as E1′-M1-M1 show the same diagram but now with M1 and M1 as the operators in the primary transitions.

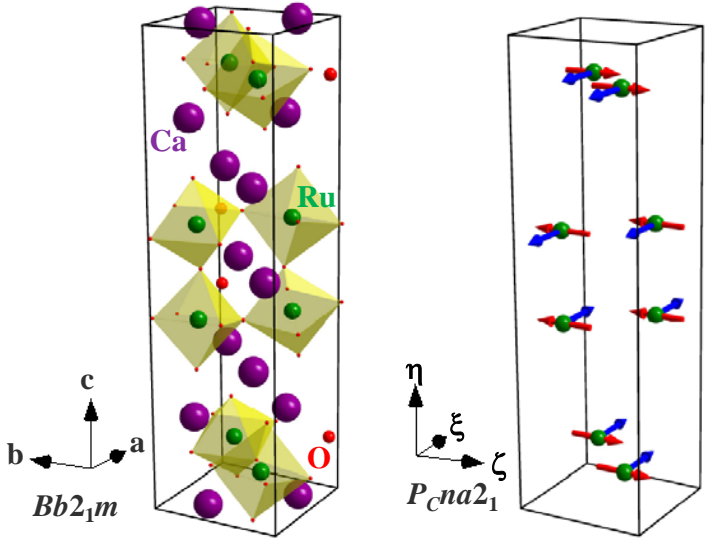

**FIG. 2.** Crystal and magnetic structures of $Ca_3Ru_2O_7$ at a temperature below 48 K. *A*-type antiferromagnetic order of axial dipoles and a propagation vector along the *c* axis. The $RuO_2$ bilayers are ferromagnetically ordered in the *a-b* plane and stacked normal to the plane. Axes (ξ, η, ζ) ≡ (*a, c, −b*) for the magnetic structure $P_Cna2_1$ are displayed. Anapoles parallel to the crystal *a* axis are depicted together with principal axial dipoles parallel to the crystal *b* axis.

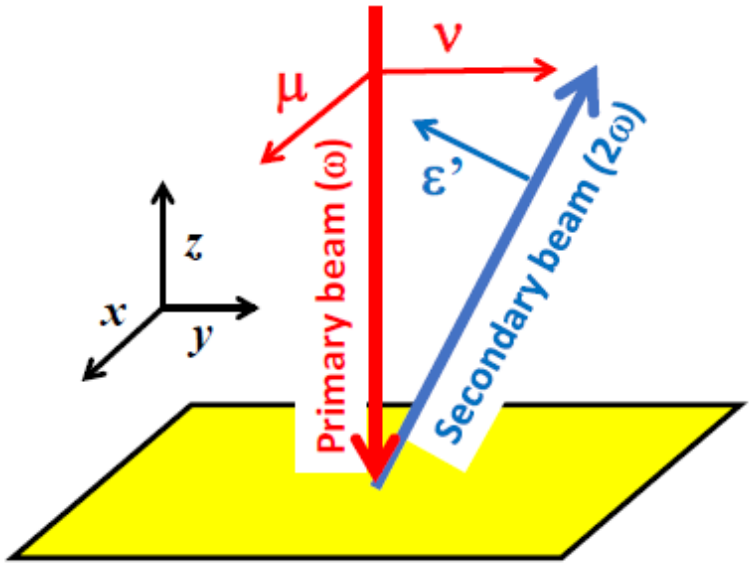

**FIG. 3.** Schematic view of experimental geometry used here for SHG. Primary polarization vectors are **μ** = (1, 0, 0) or **ν** = (0, 1, 0). The secondary beam is inclined to the *z*-axis and polarization vectors **ε'** include a component (0, 0, 1).